# From Koopman–von Neumann theory to quantum theory

## U. Klein

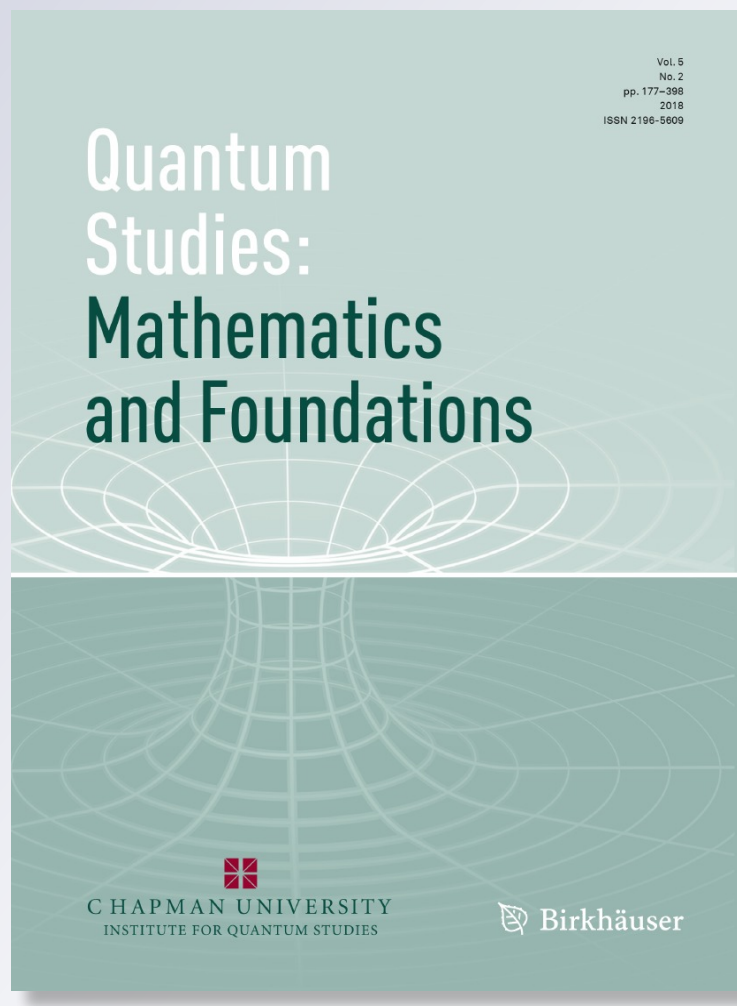

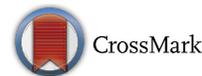

REGULAR PAPER

# From Koopman–von Neumann theory to quantum theory

U. Klein

**Abstract** Koopman and von Neumann (KvN) extended the Liouville equation by introducing a phase space function $S^{(K)}(q, p, t)$ whose physical meaning is unknown. We show that a different $S(q, p, t)$, with well-defined physical meaning, may be introduced without destroying the attractive "quantum-like" mathematical features of the KvN theory. This new $S(q, p, t)$ is the classical action expressed in phase space coordinates. It defines a mapping between observables and operators which preserves the Lie bracket structure. The new evolution equation reduces to Schrödinger's equation if functions on phase space are reduced to functions on configuration space. This new kind of "quantization" does not only establish a correspondence between observables and operators, but provides in addition a *derivation* of quantum operators and evolution equations from corresponding classical entities. It is performed by replacing $\frac{\partial}{\partial p}$ by 0 and $p$ by $\frac{\hbar}{\imath}\frac{\partial}{\partial q}$, thus providing an explanation for the common quantization rules.

## 1 Introduction

There are two fundamentally different opinions as regards the interpretation of the formalism of quantum theory (QT). These two opinions may be associated with the names of Albert Einstein and Niels Bohr. The first opinion, favored by Einstein (and in fact also by the early Bohr), says that QT is a theory about statistical ensembles. This means, the predictions of QT must be tested experimentally by performing a large number of experiments on identically prepared individual systems (statistical ensembles). The second opinion, favored by Bohr, says that QT makes predictions about individual systems. In this note, we use as a starting point a theory which can be associated with the first line of reasoning, namely the quantum-like version of the probabilistic description of classical particles due to Koopman and von Neumann [1,2].

U. Klein (✉)
Institute for Theoretical Physics, University of Linz, 4040 Linz, Austria
e-mail: ulf.klein@jku.at

A probabilistic ensemble in $2n$-dimensional phase space, with probability density $\rho(q, p, t)$, is described by the Liouville equation. To make this equation look more "quantum-like" we multiply it with a factor $\frac{\hbar}{\iota}$ and write it in the form

$$\left(\frac{\hbar}{\iota}\frac{\partial}{\partial t} + \hat{K}\right)\rho = 0, \tag{1}$$

$$\hat{K} = -\frac{\hbar}{\iota}\hat{D}_H = -\frac{\hbar}{\iota}\sum_{k=1}^{n}\left(\frac{\partial H}{\partial q_k}\frac{\partial \cdot}{\partial p_k} - \frac{\partial H}{\partial p_k}\frac{\partial \cdot}{\partial q_k}\right). \tag{2}$$

The linear operator $\hat{D}_H$ is the Lie derivative defined by $H$. In terms of the usual notation $\{A, B\}$, for the Poisson bracket of two observables $A$, $B$, it may also be written in the form $\hat{D}_H = \{H, \cdot\}$. The extraordinary simple structure of the Liouville equation leads to a enormous manifold of solutions: Given two solutions $g, h$ their product $gh$, as well as their sum, is again a solution; given a solution $\rho$, an arbitrary function of $\rho$ is again a solution. As a consequence (1) may be rewritten in the form

$$\left(\frac{\hbar}{\iota}\frac{\partial}{\partial t} + \hat{K}\right)\psi = 0, \tag{3}$$

where $\psi$ is a complex-valued variable, whose amplitude and phase (or real and imaginary parts) both obey the original Eq. (1). The new complex variable may be written in many different ways as a function of $\rho$; a most useful choice is

$$\psi = \sqrt{\rho}\exp\left[\frac{\iota}{\hbar}S^{(K)}(q, p, t)\right]. \tag{4}$$

Introducing a new dynamic variable $S^{(K)}(q, p, t)$ makes the classical theory more similar, at least formally, to QT. Continuing along this lines, we define an inner product

$$(\phi, \psi) = \int dq dp\, \phi^*(q, p)\psi(q, p). \tag{5}$$

It is now easy to see that the operator $\hat{K}$ is self-adjoint with regard to this inner product. The relations (2)–(5) represent the basic equations of the Koopman–von Neumann (KvN) theory [1,2], as developed by Gozzi, Mauro, Pagani [3–5] and others.

## 2 Discussion of the Koopman–von Neumann theory

According to its definition the differential equation (3) is equivalent to two decoupled Liouville equations for $\rho$ (or $\sqrt{\rho}$) and $S^{(K)}$. The physical (probabilistic) predictions of the KvN equation are, therefore, exactly the same as the predictions of the original Liouville equation, as shown explicitly in Ref. [3]. The results obtained for $S^{(K)}$ are irrelevant (This is, in fact, a consistency requirement because we do not know what $S^{(K)}$ means). The advantage of (3) is that it offers a variety of mathematical methods, know from QT, for the classical domain; several examples have been reported in the literature [3–5].

Let us next ask if the KvN theory is the only possibility to extend the Liouville equation in this purely technical sense (we exclude for the moment questions as for the physical meaning of $S$). Any "quantum-like", but still fundamentally classical, extension of (1) must obey the following conditions

- Linearity of the evolution equation for $\psi$,
- Self-adjointness of the relevant differential operator,

– Decoupled differential equations for $\sqrt{\rho}$ and $S$.

The last condition could be called the "classicality condition". It means that the laws of probability theory do not interfere with the deterministic laws describing the system. This is, simple as it is, a most fundamental characterization of classical probabilistic physics.

Looking for the simplest modifications of (3), one finds that addition of an arbitrary phase space function, say $-\bar{L}_H$, to $\hat{K}$ is compatible with all three requirements. Thus, we perform the replacement

$$\hat{K} \Rightarrow \hat{L}_H = -\frac{\hbar}{\imath}\hat{D}_H - \bar{L}_H. \tag{6}$$

Here, the arbitrary (and still unknown) function $\bar{L}(q, p, t)$ has been labeled by an index $H$, taking the possibility into account that $\hat{L}_H$ is a generalized Lie derivative with respect to $H$. The basic Eq. (3) is replaced by

$$\left(\frac{\hbar}{\imath}\frac{\partial}{\partial t} + \hat{L}_H\right)\psi = 0, \tag{7}$$

We can easily see that all three conditions are fulfilled. Equation (7) is obviously linear. A short calculation shows that $\hat{L}_H$ is self-adjoint. Finally, inserting (4) (with $S^{(K)}$ replaced by $S$) in (7) and separating real- and imaginary parts we obtain the two decoupled equations

$$\left[\frac{\partial}{\partial t} - \hat{D}_H\right]\sqrt{\rho} = 0, \tag{8}$$

$$\left[\frac{\partial}{\partial t} - \hat{D}_H\right]S = \bar{L}_H. \tag{9}$$

Equations (8) and (9) differ from the corresponding result of the KvN equation (3) by the inhomogeneous term $\bar{L}_H$ in Eq. (9). If $\sqrt{\rho}$ and $S$ are, however, combined according to (4), the resulting Eq. (7) has no inhomogeneous term anymore and a "quantum-like" unrestricted superposition principle holds true.

Thus, there is an infinite number of classical "quantum-like" versions of the Liouville equation, each one characterized by a particular phase space function $\bar{L}_H$. These theories are all equivalent, as regards their physical content. On the other hand they differ with regard to the details of the "quantum-like" mathematics they offer, because the self-adjoint operators $\hat{L}_H$ are different. The KvN theory, obtained by setting $\bar{L}_H = 0$, is just one of these versions.

## 3 A new extension of Liouville's equation

This variety of theories leads of course to the question which $\bar{L}_H$ is "the best". Are the $\bar{L}_H$ really all equivalent in a physical sense ? This is certainly true, as long as we stay strictly within the realm of classical physics. It is, however, not true if we widen our horizon and take the relation between classical physics and QT into account. A proper choice of $\bar{L}_H$ could possibly close the gap between classical physics and QT or help, at least, to elucidate the still rather mysterious relation between these two fields.

Clearly, a proper choice of $\bar{L}_H$ has to ascribe, first of all, a clear physical meaning to the variable $S$. Let us consider the canonical equations for a Hamiltonian $H(q, p)$, not depending on time,

$$\dot{q}_k = \frac{\partial H(q, p)}{\partial p_k}, \quad \dot{p}_k = -\frac{\partial H(q, p)}{\partial q_k}. \tag{10}$$

The solutions of (10) for initial values $q_s$, $p_s$ at time $t_s$ are written (omitting indices and summation signs; i.e., $qp = \sum_{i=1}^{n} q_i p_i$, etc) as

$$q = Q(q_s, p_s, t), \quad p = P(q_s, p_s, t). \tag{11}$$

These relations define, for each value of $t$, a mapping of phase space onto itself, from initial values $q_s$, $p_s$ at $t_s$ to the values $q$, $p$ at some later time $t$. This mapping is also referred to as the *flow* defined by $H$.

As it is well-known, the evolution in time, described by the mapping (11) may be interpreted as a canonical transformation [6,7]. As a consequence a generating function $S_H(q_s, p_s, t)$ must exist. As shown in advanced text books [6,7], it takes the form

$$S_H(q_s, p_s, t) = \int_{t_s}^{t} du \left[ P(q_s, p_s, u) \frac{\partial Q(q_s, p_s, u)}{\partial u} - H(Q(q_s, p_s, u), P(q_s, p_s, u)) \right], \quad (12)$$

The integrand has the form of a Lagrangian $L_H$, associated with $H$, and $S_H$ is the corresponding classical action. The latter is written here as a function of initial values. Generally, the action (as a functional evaluated at the stationary points) depends on $2n + 2$ variables, namely $2n$ integration constants and two boundary points. In the most common representation the integration constants are identified with the coordinates of the initial and final positions, $q_s$ and $q$, in the present representation they are identified with the initial positions and momenta $q_s$, $p_s$.

The physical meaning of Eq. (12) is that of an integrability condition for the canonical equations (10). Thus, relation (12), defining $S_H$, must be considered as an integral part of the present theory; this makes it a natural candidate for our field $S$. Let us investigate if indeed an associated phase space function $S_H(q, p, t)$ exists, which fulfills Eq. (9). As the solutions (11) represent a canonical transformation, they may be inverted to give the initial values $q_s$, $p_s$ as a function of $q$, $p$, $t$,

$$q_s = Q_s(q, p, t), \quad p_s = P_s(q, p, t). \quad (13)$$

Let us consider an arbitrary function $G_s$ which depends, like our $S_H$, on the initial values $q_s$, $p_s$ and also explicitly on time $t$. Using (13) we obtain for each $G_s(q_s, p_s, t)$ an associated phase space function $G(q, p, t)$, which is defined by

$$G_s(Q_s(q, p, t), P_s(q, p, t), t) = G(q, p, t). \quad (14)$$

The function $G$ is, in principle, determined by the form of $G_s$ and by the solutions of the equations of motion. Alternatively, a partial differential equation for $G$ may be derived [8]. It may be found using the fact that the initial values are "constants of motion". Calculating the derivative of (14) with respect to $t$, and replacing $\dot{q}$ and $\dot{p}$ by phase space functions according to (10), we obtain the differential equation

$$\frac{\partial G}{\partial t} + \frac{\partial G}{\partial q} \frac{\partial H}{\partial p} - \frac{\partial G}{\partial p} \frac{\partial H}{\partial q} = \frac{\partial G_s}{\partial t}. \quad (15)$$

If $G$ is identified with $S_H$, this equation takes indeed the form of (9), where $\bar{L}_H$ is given by the Lagrangian,

$$\left[ \frac{\partial}{\partial t} - \hat{D}_H \right] S_H = \bar{L}_H, \quad (16)$$

$$\bar{L}_H(q, p) = p \frac{\partial H(q, p)}{\partial p} - H(q, p). \quad (17)$$

Note that the first of the canonical equations (10) has been used to replace $\dot{q}$ by the derivative of $H$ with respect to $p$. We arrived at a new extension of Liouville's equation which takes the form (7), with a differential operator $\hat{L}_H$ defined by (6) and (17), and a "classical wave function"

$$\psi(q, p, t) = \sqrt{\rho(q, p, t)} \exp\left[ \frac{\iota}{\hbar} S(q, p, t) \right], \quad (18)$$

with a well-defined physical meaning; its squared modulus is the classical probability density $\rho(q, p, t)$ and its phase (we omit the index $H$) is the classical action $S(q, p, t)$. As mentioned already this is an entirely classical equation; the dependence on $\hbar$ is spurious.

## 4 Projection of the new equation to configuration space

As noted already, all possible extensions of Liouville's equation, following the rule (6), are equivalent in a purely classical sense. The question is, if the new equation is able to contribute to our understanding of the intricate relation between classical physics and QT. If the definition of $\hat{L}_H$ is inserted, Eq. (7) takes the form

$$\left[\frac{\hbar}{\iota}\frac{\partial}{\partial t} - \frac{\hbar}{\iota}\frac{\partial H}{\partial q_k}\frac{\partial}{\partial p_k} + \frac{\partial H}{\partial p_k}\left(\frac{\hbar}{\iota}\frac{\partial}{\partial q_k} - p_k\right) + H\right]\psi^{(c)} = 0. \tag{19}$$

An obvious but very important difference between the classical wave function, denoted here for distinction by $\psi^{(c)}$, and the true quantum wave function, denoted by $\psi$, is the number of independent variables; $\psi^{(c)}$ depends, neglecting time for the moment, on $2n$ variables (typically the standard phase space variables $q$, $p$ used in this paper) while $\psi$ depends on $n$ variables (typically the configuration space variables $q$). A quantum state is a point in Hilbert space, or a function defined on configuration space. There is general agreement with regard to this statement. On the other hand, there is no agreement as regards the question what to consider as classical counterpart of a quantum state. It is widely believed, following Bohr, that the classical counterpart of a quantum state is a point in phase space, describing the "position" of a single mechanical system. Here, we assume, following Einstein, that the classical counterpart of a quantum state is a classical probabilistic ensemble, described by a probability density $\rho(q, p)$, or a point $\psi^{(c)}$ in an associated classical Hilbert space.

Considered from this point of view, which will be further confirmed elsewhere, the transition to QT is basically a kind of projection from phase space to configuration space, with an associated reduction in the number of degrees of freedom. This process may be performed in several ways. The way of proceeding reported in this paper is not the most complete one, but seems attractive because of its extraordinary simplicity and its ability to explain the quantization rules in an intuitive way. The classical law in phase space (19) must be projected to a quantum law in configuration space, with states $\psi$ depending only on $q$, and a differential operator containing only derivatives with respect to $q$. This can obviously be done by implementing the rules

$$\frac{\partial}{\partial p_k} = 0, \quad p_k = \frac{\hbar}{\iota}\frac{\partial}{\partial q_k} \tag{20}$$

in the differential operator acting on $\psi^{(c)}$. The left member of (20) does not (explicitly) belong to the standard quantization rules; it simply expresses the fact that the new states do not depend on the generalized coordinates $p$. The right member of (20) is the standard quantization rule, expressing the strange fact that a variable must be replaced by an operator. Implementing these rules in Eq. (19) creates the famous differential equation

$$\left[\frac{\hbar}{\iota}\frac{\partial}{\partial t} + H\left(q, \frac{\hbar}{\iota}\frac{\partial}{\partial q_k}\right)\right]\psi(q, t) = 0, \tag{21}$$

found by Schrödinger. Of course, the parameter $\hbar$ can now, as a consequence of the modification of the original equation, no longer be eliminated, but becomes instead a new fundamental constant. This derivation shows that the version (19) of the Liouville equation may be interpreted as classical counterpart of the Schrödinger equation. The common quantization rule, replacing $p$ by a derivative with respect to $q$, can be understood as part of a projection from phase space to configuration space.

Note that, in contrast to the common treatment, the present "quantization" starts from a classical evolution equation and not from a classical Hamiltonian function. The process of quantization creates in a single step both the new evolution equation and the new Hamiltonian operator.

We should note at this point that the form of $\bar{L}_H(q, p)$ given in (17) is not the most general one. An infinite number of admissible $\bar{L}_H(q, p)$ may be found by adding to (17) a total derivative of an arbitrary function $C(q, p)$; this kind of gauge freedom in the Lagrangian is actually well-known. We restricted ourselves in the present paper to the choice $C(q, p) = 0$ associated with the most common (configuration space) representation of the Schrödinger equation. Other choices lead to the Schrödinger equation in momentum space and to the "phase-space Schrödinger equation" derived by Torres-Vega and Frederick [9], as will be shown in more detail in later work.

## 5 The correspondence between observables and operators

The above development [see Eqs. (6), (17)] shows that the operator $\hat{L}_H$ is, in the course of the projection from phase space to configuration space, reduced to the Hamiltonian operator $\hat{H}$,

$$\hat{L}_H \searrow \hat{H} = H\left(q, \frac{\hbar}{\iota} \frac{\partial}{\partial q_k}\right). \tag{22}$$

The operators $\hat{L}_H$ and $\hat{H}$ are completely determined by the analytical form of the classical phase space function $H(q, p)$; we neglect ordering problems, which actually do not play any role for the most important observables. The above rules for the assignment of phase space operators to classical observables and subsequent reduction to configuration space operators, may therefore, be applied to arbitrary observables $A(q, p)$:

$$A(q, p) \Rightarrow \hat{L}_A = -\frac{\hbar}{\iota} \hat{D}_A - \bar{L}_A = A - \frac{\partial A}{\partial p_k}\left(p_k - \frac{\hbar}{\iota} \frac{\partial}{\partial q_k}\right) - \frac{\hbar}{\iota} \frac{\partial A}{\partial q_k} \frac{\partial}{\partial p_k} \searrow \hat{A} = A\left(q, \frac{\hbar}{\iota} \frac{\partial}{\partial q_k}\right). \tag{23}$$

The two processes displayed in Eq. (23) will be abbreviated by the statement $A \to \hat{A}$. Considered from the point of view of probability theory, the fact that observables become operators represents a breakdown of the standard concept of random variables.

From a formal point of view, the above quantization process consists of two consecutive steps. In a first step, abbreviated here as $\Rightarrow$, phase space operators are defined. In a second step, abbreviated as $\searrow$, configuration (or momentum) space operators are created. After completion of this work, the author learned that both of these steps occur also in a mathematical discipline called geometric (de)quantization [10–12] and are denoted there as "prequantization" and "polarization". Unfortunately, the physical content of most of these works is not easy to access due to their highly abstract and sometimes speculative character. On the other hand, it is to be expected, that this high-level mathematical work contains many important results. More popular accounts (such as section II D of [11]) would be highly desirable.

5.1 Several observables

It is clear that commutativity under ordinary multiplication of two functions $A$, $B$ is, in general, not preserved under these transitions: if $A \to \hat{A}$, $B \to \hat{B}$, then, in general, $\hat{A}\hat{B} \neq \hat{B}\hat{A}$.

The fact that the commutativity of classical observables is not recovered in the corresponding operator structure is no surprise in the present approach: A state in QT has a whole ensemble of trajectories as its classical counterpart, and the notion of classical observables, which is closely related to the idea that all properties of a system are determined by a single point in phase space, becomes meaningless. We cannot expect that structural properties of classical observables are mapped in a one-to-one manner to corresponding properties of operators. The impossibility to achieve this, by any reordering of terms, is the content of Groenewold's theorem [13].

A "quantum-like" non-commuting structure may, in fact, already be found at the classical level, if ensembles of trajectories are taken into consideration. Each observable $A$ creates its own flow with the help of canonical equations like (10). A second observable $B$ is invariant under the flow created by $A$, if

$$\hat{D}_A B = \{A, B\} = 0. \tag{24}$$

In this case, the inverse, namely the invariance of $A$ under the flow created by $B$, is also true. Two Lie derivatives performed in consecutive order do, in general, not commute. It can, however, be shown (see, for example [6]) that

$$[\hat{D}_A, \hat{D}_B] = \hat{D}_{\{A,B\}}, \tag{25}$$

holds true, where the bracket is an abbreviation for the commutator of $\hat{D}_A$ and $\hat{D}_B$ (the latter is defined, for general operators $\hat{U}$, $\hat{V}$, by $[\hat{U}, \hat{V}] = \hat{U}\hat{V} - \hat{V}\hat{U}$). We see here a certain analogy between Poisson brackets and commutators. This leads to the question if a mapping from Poisson brackets to quantum commutators does exist. More precisely, does a mapping $A \to \hat{O}_A$ from phase space observables $A$ to quantum operators $\hat{O}_A$ exist, such that $A \to \hat{O}_A$, $B \to \hat{O}_B$, and $\{A, B\} \to \hat{O}_{\{A,B\}}$ implies $[\hat{O}_A, \hat{O}_B] = \hat{O}_{\{A,B\}}$? A further consistency requirement is the condition $\hat{O}_1 = \hat{1}$. If these conditions are fulfilled, we will say that the Lie bracket structure is preserved.

It is seen at once, that the Lie bracket structure is not preserved under the rule $A \Rightarrow \hat{D}_A$, because the condition $\hat{D}_1 = \hat{1}$ does not hold. Also, the correct position operator of QT is not obtained if the projection to configuration space is performed. Thus, the assignment used in the KvN theory, does not preserve the Lie bracket structure.

5.2 The assignment $A \Rightarrow \hat{L}_A$

Let us now study the new assignment $A \Rightarrow \hat{L}_A$, defined by (23). We will first ask if the Lie bracket structure is preserved in the transition from observables to phase space operators. In a second step, the transition to quantum operators, will be considered. The extended operator $\hat{L}_A = -\frac{\hbar}{\iota}\hat{D}_A - \bar{L}_A$ fulfills obviously the condition $\hat{L}_1 = \hat{1}$. Furthermore, $\hat{L}_A$ obeys the condition

$$[\hat{L}_A, \hat{L}_B] = -\frac{\hbar}{\iota}\hat{L}_{\{A,B\}}, \tag{26}$$

which shows that the Lie bracket structure is preserved, as far as the transition to phase space operators is concerned. To prove (26) it is convenient to write $\hat{L}_A$ tentatively in the form

$$\hat{L}_A = -\frac{\hbar}{\iota}\hat{D}_A + M[A], \tag{27}$$

where $M[A]$ is a still unknown phase space function determined by $A$. Inserting this Ansatz into Eqs. (26) and using (25) leads to the condition

$$\{A, M[B]\} - \{B, M[A]\} - M[\{A, B\}] = 0, \tag{28}$$

for $M[A]$. It can now be shown by straightforward calculation that $M[A] = -\bar{L}_A$ is a solution of (28). This proves that the differential operator $\hat{L}_H$, occurring in our new Eq. (19), has been constructed by means of a "reasonable", Lie-bracket-preserving assignment $A \Rightarrow \hat{L}_A$. The present result for $\hat{L}_A$ has been derived before, in the framework of a mathematical process called "prequantization". It has been shown [14] that $\hat{L}_A$ as given by (23) is the most general solution of the problem defined by Eqs. (26), (27) and the condition $\hat{L}_1 = \hat{1}$. This result confirms our construction of the classical Schrödinger equation (19).

To perform, in a second step, the transition to QT, Eq. (26) must be projected to configuration space, using the rules (20). This projection will certainly not preserve the Lie bracket structure for *all* observables. Considering for simplicity a single particle (the case $n = 3$), the most important observables are position $q_k$, momentum $p_k$, angular momentum $l_k = \epsilon_{kil} q_i p_l$, and kinetic energy $t = \frac{p_j p_j}{2m}$.

The well-known quantum operators $\hat{q}_k$, $\hat{p}_k$, $\hat{l}_k$, $\hat{t}$ associated with these observables are obtained at once from Eq. (23) (no ordering problem here). Actually, these operators—or their higher-dimensional counterparts—are the only ones which play a role in real-world applications of QT (neglecting here spin, which will be dealt with in forthcoming work). The reason is probably that the corresponding observables describe fundamental properties of space, or may be conserved quantities as a consequence of corresponding symmetries. This defines a *context* for these operators, within which measurements may be performed in a meaningful way, even if the classical meaning of observables has been lost in QT.

Calculating the Poisson brackets of $q_k$, $p_k$, $l_k$, $t$ one obtains

$$\{q_i, q_k\} = 0, \quad \{p_i, p_k\} = 0, \quad \{q_i, p_k\} = \delta_{ik}$$
$$\{l_i, l_k\} = \epsilon_{ikl} l_l, \quad \{q_i, l_k\} = \epsilon_{ikl} q_l, \quad \{p_i, l_k\} = \epsilon_{ikl} p_l$$
$$\{q_i, t\} = \frac{p_i}{m}, \quad \{p_i, t\} = 0, \quad \{l_i, t\} = 0. \tag{29}$$

To transform this Poisson bracket structure to QT we must first assign phase space operators $\hat{L}_{q_k}$, $\hat{L}_{p_k}$, $\hat{L}_{l_k}$, $\hat{L}_t$ to $q_k$, $p_k$, $l_k$, $t$, and then, in a second step, project the corresponding commutator relations (26) to configuration space. Omitting the details of this calculation, the result is given by

$$[\hat{q}_i, \hat{q}_k] = 0, \quad [\hat{p}_i, \hat{p}_k] = 0, \quad [\hat{q}_i, \hat{p}_k] = -\frac{\hbar}{\iota}\delta_{ik}$$
$$[\hat{l}_i, \hat{l}_k] = -\frac{\hbar}{\iota}\epsilon_{ikl}\hat{l}_l, \quad [\hat{q}_i, \hat{l}_k] = -\frac{\hbar}{\iota}\epsilon_{ikl}\hat{q}_l, \quad [\hat{p}_i, \hat{l}_k] = -\frac{\hbar}{\iota}\epsilon_{ikl}\hat{p}_l$$
$$[\hat{q}_i, \hat{t}] = -\frac{\hbar}{\iota}\frac{\hat{p}_i}{m}, \quad [\hat{p}_i, \hat{t}] = 0, \quad [\hat{l}_i, \hat{t}] = 0. \tag{30}$$

This result, obtained here from classical theory, may of course be verified with the help of the explicit definitions of the operators $\hat{q}_k$, $\hat{p}_k$, $\hat{l}_k$, $\hat{t}$.

The above commutator structure may also be obtained by "translation" from the classical formulas, by means of the following translation rule: If $A \to \hat{A}$, $B \to \hat{B}$, $C \to \hat{C}$ and $C = \{A, B\}$, then

$$\hat{C} = -\frac{\iota}{\hbar}[\hat{A}, \hat{B}]. \tag{31}$$

This structural similarity represents one of the strongest links between classical physics and QT. Despite its success, nobody knows where this rule comes from. The present derivation sheds some light on this question: Apparently, quantum operators are related to infinitesimal generators of (canonical) transformations, which describe the flow created by classical observables in phase space. The quantum commutator relations are a vestige of the corresponding phase space relations (26), which itself describe, in a way not yet understood in detail, the commutativity properties of the flow. The flow describes not an individual system but a statistical ensemble.

**Acknowledgements** Open access funding provided by Johannes Kepler University Linz.